\documentclass[12pt]{article}
\usepackage[english]{babel}
\usepackage{url}
\usepackage[utf8x]{inputenc}
\usepackage{amsmath}
\usepackage{graphicx}
\graphicspath{{images/}}
\usepackage{parskip}
\usepackage{fancyhdr}
\usepackage{vmargin}
\usepackage{adjustbox}
\setmarginsrb{1.5 cm}{0.5 cm}{1.5 cm}{0.5 cm}{1 cm}{1.5 cm}{1 cm}{1.5 cm}

\title{Cerebral Signal Instantaneous Parameters Estimation MATLAB Toolbox - User Guide Version 2.3}                             
\author{Esmaeil Seraj}                               
\date{\today}                                           

\makeatletter
\let\thetitle\@title
\let\theauthor\@author
\let\thedate\@date
\makeatother

\begin{document}


\begin{titlepage}
    \centering 
	\includegraphics[scale = 0.65]{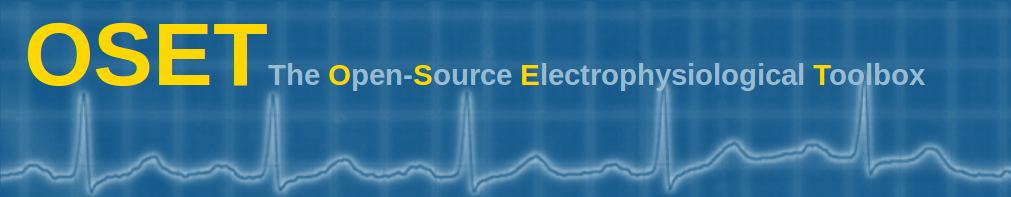}\\[1.0 cm]
    \textsc{\Large The Open-Source Electrophysiological Toolbox (OSET)}\\[1 cm]    
    \includegraphics[scale = 0.3]{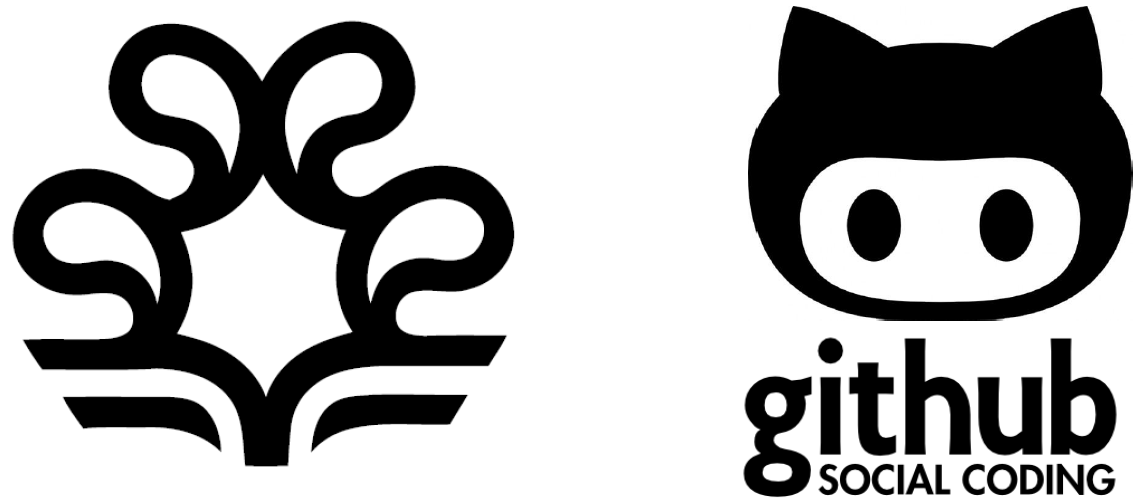}\\[1.0 cm] 
    \textsc{\Large Shiraz University, Department of Computer Science \& Information Technology}\\[0.5 cm]               
    \textsc{\large School of Electrical and Computer Engineering, Shiraz, Iran}\\[1 cm]               
    \rule{\linewidth}{0.5 mm} \\[0.4 cm]
    { \Large \bfseries \thetitle}\\
    \rule{\linewidth}{0.5 mm} \\[2.5 cm]
    
    \begin{minipage}{0.4\textwidth}
        \begin{flushleft} \large
            \textbf{Provider:}\\
            \theauthor\\
            esmaeil.seraj09@gmail.com
            \end{flushleft}
            \end{minipage}~
            \begin{minipage}{0.4\textwidth}
            \begin{flushright} \large
            \textbf{Supervisor:} \\
             Dr. R.Sameni\\
             reza.sameni@gmail.com             
        \end{flushright}
    \end{minipage}\\[2 cm]
    
    {\large \thedate}\\[2 cm]
 
    \vfill
    
\end{titlepage}


\vspace*{5 cm} 
\begin{abstract}
\noindent This document is meant to help individuals use the Cerebral Signal Instantaneous Parameters Estimation MATLAB Toolbox which implements different methods for estimating the instantaneous parameters of cerebral signals (i.e. EEG, MEG and etc. ) such as phase, frequency and envelope of a signal and calculating some related well-known quantities in a variety of applications.

\noindent The toolbox – which is distributed under the terms of the GNU GENERAL PUBLIC LICENSE as a set of MATLAB\textregistered ~routines – can be downloaded directly at the address:

 \begin{center}
 \underline{\texttt{http://oset.ir/category.php?dir=Tools}}
  \end{center}.
 
 or from the public repository on GitHub, at address below:

 \begin{center} 
 \underline{\texttt{https://github.com/EsiSeraj/EEG-PhaseFreq-Analysis}}
 \end{center}.

\noindent The purpose of this toolbox is to calculate the instantaneous phase, frequency and envelope sequences of cerebral signals (EEG, MEG, etc.) and some related popular features and quantities in brain studies and Neuroscience such as Phase Synchronization, Frequency Shift, Phase Resetting, Phase Locking Value (PLV), Phase Difference and more, to help researchers in these fields. 
\end{abstract}


\newpage
\tableofcontents
\pagebreak


\section{Getting Started}
\label{sec:gettingstarted}

This document is meant to help individuals use the Cerebral Signal Instantaneous Parameters Estimation MATLAB Toolbox which implements different methods, including recently proposed Transfer Function Perturbation (TFP) framework through both procedures introduced in \cite{ESeraj1} and \cite{ESeraj2}, for estimating the instantaneous parameters of cerebral signals (i.e. EEG, MEG and etc. ) such as phase, frequency and envelope of a signal and calculating some related well-known quantities in a variety of applications, as introduced and utilized in different studies such as \cite{ESeraj5, ESeraj6, ESeraj7} and  \cite{ESeraj8}. The references and utilized test data are introduced later.

\subsection{License - No Warranty}
\label{subsec:license}
This program is free software; you can redistribute it and/or modify it under the terms of the GNU GENERAL PUBLIC LICENSE as published by the Free Software Foundation; either version 2 of the License, or (at your option) any later version.
This program is distributed in the hope that it will be useful, but WITHOUT ANY WARRANTY; without even the implied warranty of MERCHANTABILITY or FITNESS FOR A PARTICULAR PURPOSE. See the GNU GENERAL PUBLIC LICENSE for more details. You should have received a copy of the GNU GENERAL PUBLIC LICENSE along with this program; if not, write to the Free Software Foundation, Inc., 51 Franklin Street, Fifth Floor, Boston, MA  02110-1301, USA.

\subsubsection{The Open-Source Electrophysiological Toolbox (OSET)}
\label{subsub:oset}
Open Source Electrophysiological Toolbox (OSET) is a collection of electrophysiological data and open source codes for biosignal generation, modeling, processing, and filtering. OSET, version 3.1, 2014 Released under the GNU GENERAL PUBLIC LICENSE. Copyright\textcopyright ~2012.

Reza Sameni (reza.sameni@gmail.com), Department of Computer Science \& Information Technology, School of Electrical and Computer Engineering, Shiraz University, Shiraz, Iran.

\subsection{Citation}
\label{subsec:citation}
Within the limits of the GNU GENERAL PUBLIC LICENSE, you can use the toolbox as you please. If you use the toolbox in a work of your own that you wish to publish, please make sure to cite this user manual \cite{ESeraj3},  the original studies \cite{ESeraj1, ESeraj2} and the OSET \cite{RSameni} properly, as shown below. This way you will contribute to helping other scholars find these items.

\begin{itemize}

\item Esmaeil Seraj, “Cerebral Signal Instantaneous Parameters Estimation MATLAB Toolbox - User Guide Version 2.3,” arXiv Preprint, Dec. 2017 [Online]. Available: \underline{https://arxiv.org/abs/1610.02249}

\item Esmaeil Seraj and Reza Sameni, “Robust Electroencephalogram Phase Estimation with Applications in Brain-computer Interface Systems,”Physiological Measurement, vol. 38, no. 3, pp. 501–523, Feb. 2017. [Published Online]DOI: 10.1088/1361-
6579/aa5bba

\item  Reza Sameni and Esmaeil Seraj, “A Robust Statistical Framework for Instantaneous Electroencephalogram Phase and Frequency Analysis,” Physiological Measurement, vol. 38, no. 12, pp. 2141–2163, Nov. 2017. [Published Online] DOI:
10.1088/1361-6579/aa93a1

\item Reza Sameni, The Open-Source Electrophysiological Toolbox (OSET), [Online] version 3.1 (2014). URL \underline{http://www.oset.ir}
\end{itemize}

\subsection{Download and Utilization}
\label{subsec:download}
The latest version of the toolbox can be downloaded at:
\begin{center}
{\texttt{\underline{http://oset.ir/category.php?dir=Tools}}}
\end{center} 

The functions and m-files can be downloaded separately as you need or all together in a compressed file named \texttt{Phase\_Analysis\_Toolbox}. Once you have downloaded and uncompressed the toolbox, 36 files represented in following Tables shall be appeared. Additionally, you might add the path of the directory in which you stored the toolbox to your MATLAB in order to easily use and apply the functions for your own dataset.

\begin{table}[h!]
\centering
\label{tab:data&manual}
\caption{Sample Data and User Manual}
\begin{tabular}{ |c|c||c|c| }
\hline
row & name & row & name \\ 
 \hline
 \hline
 1 & LICENSE.txt & 6 & User\_Guide\_Ver2.3.pdf \\
 \hline 
 2 & \tt{EEG.m} & 7 & \tt{Z\_phase.mat} \\ 
 \hline
 3 & \tt{Z001\_phase.mat} & 8 & \tt{Z001.txt} \\ 
 \hline
 4 & \tt{Z002\_phase.mat} & 9 & \tt{Z002.txt} \\
 \hline 
 5 & \tt{Z003\_phase.mat} & 10 & \tt{Z003.txt} \\ 
 \hline
\end{tabular}
\end{table}

\begin{table}[h!]
\centering
\label{tab:ZPPPmfiles}
\caption{Conventional, Robust Phase Calculation \cite{ESeraj1} and \cite{ESeraj2} Methods Alongside with Phase Feature Estimators From \cite{ESeraj5, ESeraj6, ESeraj7} and  \cite{ESeraj8}}.
\begin{tabular}{ |c|c||c|c| }
\hline
row & name (function) & row & name (test \tt{m-files}) \\ 
 \hline
 \hline
 1 & \tt{Phase\_Ex\_TFP.m} & 10 & \tt{Test\_Phase\_Ex\_TFP.m} \\
 \hline
 2 & \tt{Phase\_Ex\_ZPPP.m} & 11 & \tt{Test\_Phase\_Ex\_ZPPP.m} \\
 \hline 
 3 & \tt{Phase\_Ex\_Trad.m} & 12 & \tt{Test\_Phase\_Ex\_Trad.m} \\ 
 \hline
 4 & \tt{Phase\_Features\_MultiCh.m} & 13 & \tt{Test\_Phase\_Features\_MultiCh.m} \\ 
 \hline
 5 & \tt{Phase\_Features\_SingleCh.m} & 14 & \tt{Test\_Phase\_Features\_SingleCh.m} \\
 \hline 
 6 & \tt{Phase\_Features.m} & 15 & \tt{Test\_Phase\_Features.m} \\ 
 \hline
 7 & \tt{PLV\_PhaseSeq.m} & 16 & \tt{Test\_PLV\_PhaseSeq.m} \\
 \hline
 8 & \tt{PLV\_RawSig.m} & 17 & \tt{Test\_PLV\_RawSig.m} \\
 \hline
 9 & \tt{Synth\_EEG.m} & 18 & \tt{Test\_Synth\_EEG.m} \\
 \hline
\end{tabular}
\end{table}

\begin{table}[h!]
\centering
\label{tab:statisticalframework}
\caption{Source Codes of Statistical Framework for EEG Phase and Frequency Analysis Represented in \cite{ESeraj2} and Required Extra Filtering Functions from OSET \cite{RSameni}}
\begin{tabular}{ |c|c||c|c| }
\hline
row & name & row & name \\ 
 \hline
 \hline
 1 & \tt{Test\_StstclFrmwrk\_SNRver\_KalmanSmooth.m} & 5 & \tt{BPFilter5.m} \\
 \hline 
 2 & \tt{Test\_StstclFrmwrk\_AmpPhase\_Distributions.m} & 6 & \tt{LPFilter.m} \\ 
 \hline
 3 & \tt{Test\_StstclFrmwrk\_Example\_LowAmp.m} & 7 & \tt{KFNotch.m} \\ 
 \hline
 4 & \tt{Test\_StstclFrmwrk\_ROCCurves.m} & 8 & \tt{KalmanSmoother.m} \\ 
 \hline
\end{tabular}
\end{table}

The file \textit{User\_Guide\_Ver2.3.pdf}, is similar to current document and contains the latest version of the cerebral signal phase analysis toolbox user manual. The rest of the files are categorized into three groups according to the tables. The main data files \textbf{Z001}, \textbf{Z002} and \textbf{Z003} presented in Table~\ref{tab:data&manual} are ongoing EEG signals employed from the dataset introduced in \cite{Andrzejak} and their corresponding instantaneous phase signals are extracted using the functions in current toolbox.

\subsection{Getting Help}
\label{subsec:help}
If you have added the toolbox directory to the MATLAB\textregistered ~path you can simply type:
 \begin{center}
\texttt{<doc~~~function\_name>} \hspace*{1cm} or \hspace*{1cm} \texttt{<help~~~function\_name>}
 \end{center} 
 
in command window to get online help for the function you are using. Furthermore, you can also contact the providers\footnote{Esmaeil Seraj: (esmaeil.seraj09@gmail.com)~~~or~~~Reza Sameni: (reza.sameni@gmail.com)} directly to ask any related questions or discuss possible difficulties or errors you might encounter. Please feel free to contact in either cases.

\section{User Guide}
\label{sec:userguide}
\subsection{Context}
\label{subsec:context}
The purpose of this toolbox is to calculate the instantaneous phase and frequency sequences of cerebral signals (EEG, MEG, etc.) and some related popular features and quantities in brain studies and Neuroscience such as Phase Shift, Phase Resetting, Phase Locking Value (PLV), Phase Difference and more, to help researchers in these fields. The original research papers introducing the estimation methods and discussing the underlaying mathematical logics are \cite{ESeraj2} and \cite{ESeraj1}. For proof of concept, these methods and feature estimation functions have been used and tested in different applications such as BCI \cite{ESeraj1, ESeraj7}, sleep stage scoring \cite{ESeraj5} and general brain synchrony and connectivity \cite{ESeraj2, ESeraj6, ESeraj8}.

\subsection{Fundamentals}
\label{subsec:principals}
\subsubsection{Conventional Instantaneous Phase Estimation Procedure}
\label{subsec:traditionalmethod}
The most common definition of the instantaneous phase is based on the analytic representation of a signal \cite{MChavez}. Unless the signal has a very narrow-band spectral support, the extracted phase signal might be meaningless \cite{MChavez}. Accordingly, for the signal $x(t)$, the first step is the frequency filtering and make the signal narrow-band $ x_f(t) $. Thereby, the analytical form of filtered signal is defined as follows \cite{BPicinbono}:
\begin{equation}
\label{eq:analytic}
z_x(t) = x_f(t) + jH\{x_f(t)\}
\end{equation}
where $H\{x_f(t)\}$ is the Hilbert transform of filtered signal. Using the analytical form, the instantaneous amplitude and phase pair are uniquely defined as follows:
\begin{equation}
a_x(t)=|z_x(t)|=\sqrt{x_f(t)^2 + H\{x_f(t)\}^2}
\label{eq:amp}
\end{equation}
\begin{equation}
\phi_x(t)=\arctan\left(\frac{H\{x_f(t)\}}{x_f(t)}\right)
\label{eq:phase}
\end{equation}

In traditional phase estimation procedure, linear-phase filters, i.e. FIR filters, are widely used through the Hilbert Transform to have the minimum distortion on the phase of signal. In this toolbox, the function \texttt{Phase\_Ex\_Trad.m} is provided for this purpose. To operate this function you can either use the test m-file \texttt{Test\_Phase\_Ex\_Trad.m} or just simply type:
\begin{center}
\texttt{[phase, inst\_freq, amp] = Phase\_Ex\_Trad(x1, fs, WS, max\_f, ndft)}
\end{center}
where the inputs and outputs are described later.

\subsubsection{Robust Monte Carlo Phase Estimation Approach (Random Perturbations in Filters Response)}
\label{subsec:zpppmethod}
Zero-pole Perturbation Phase estimation procedure (Z-PPP) is a new robust method presented for extracting the instantaneous phase of a signal \cite{ESeraj1}. This method is based on using the IIR filters in a zero-phase filtering process with absolutely no phase distortion for making the signal narrow-band and then perturbing the frequency response of the utilized filter to reduce the side-effects of the procedure. A complete description of the Z-PPP method and confronting problems and side-effects in phase estimation process is presented in \cite{ESeraj1}. In fact, in Z-PPP method, the procedure described in \ref{subsec:traditionalmethod} is modified in way that the estimated instantaneous phase signal is more reliable and meaningful.

In this toolbox, the function \texttt{Phase\_Ex\_ZPPP.m} is provided for this purpose. To operate this function you can either use the test m-file \texttt{Test\_Phase\_Ex\_ZPPP.m} or just simply type:
\begin{center}
\texttt{[phase, inst\_freq, amp] = Phase\_Ex\_ZPPP(x1, fs, WS, max\_f, ndft, pertnum)}
\end{center}
where the inputs and outputs are described later.

\subsubsection{Robust Statistical Framework for Instantaneous EEG Phase and Frequency Analysis}
\label{subsec:statisticalframework}
In our recent study \cite{ESeraj2}, we have presented a statistical framework for EEG phase analysis. Using an additive data model between the so called background (spontaneous) and foreground EEG, probability density functions and other statistical properties of the instantaneous EEG envelope, phase and frequency were derived. It was analytically and numerically shown that in low analytical signal envelopes, the EEG phase is highly noisy and susceptible to the background EEG activity. It was shown that although EEG phase variations convey important information regarding the EEG, some instantaneous phase jumps are systematic side effects of the processing stages used for EEG phase extraction in low analytical envelopes and are not related to the brain \cite{ESeraj2}.

In this research \cite{ESeraj2}, using this widely accepted phase calculation approach and well-established methods from statistical signal processing, a stochastic model is proposed for the superposition of narrow-band foreground and background EEG activities. Using this model, the probability density functions of the instantaneous envelope (IE) and IP of EEG signals are derived analytically. Based on the findings reported in \cite{ESeraj2}, we have proposed a Monte Carlo estimation scheme is for the accurate calculation and smoothing of phase related parameters. The impact of this approach on previous studies including time-domain phase synchrony, phase resetting, phase locking value and phase amplitude coupling are also studied in \cite{ESeraj2} with examples.

In this toolbox, the function \texttt{Phase\_Ex\_TFP.m} is provided for this purpose. To operate this function you can either use the test m-file \texttt{Test\_Phase\_Ex\_TFP.m} or just simply type:
\begin{center}
\texttt{[phase, inst\_freq, amp] = Phase\_Ex\_TFP(x1, fs, f0, bw\_base, bw\_base\_dev, f0\_dev, dither\_std, pertnum)}
\end{center}
where the inputs and outputs are described later.

In order to make the results of our recent study "\textbf{A Robust Statistical Framework for Instantaneous Electroencephalogram Phase and Frequency Analysis}" \cite{ESeraj2} reproducible, all source codes of this study are also available in this toolbox. For any further details, please refer to the original research \cite{ESeraj2}. Utilizing this programs requires four extra filtering functions, namely \texttt{BPFilter5.m}, \texttt{KFNotch.m}, \texttt{LPFilter.m} and \texttt{KalmanSmoother.m} from "\textit{General filtering and processing tools}" of OSET which for simplicity are included within current toolbox as well. Each \texttt{m-file} is shortly described below.

\begin{itemize}
\item The program \texttt{Test\_StstclFrmwrk\_SNRver\_KalmanSmooth.m} implements the presented \textit{robust statistical framework for instantaneous EEG phase and frequency analysis}, the SNR verification described within the paper \cite{ESeraj2} and the proposed post-processing step through Kalman Smoothing, all in once.

\item The program \texttt{Test\_StstclFrmwrk\_AmpPhase\_Distributions.m} implements and illustrates the \textit{probability density functions} (PDF) of envelope and phase, introduced in the paper \cite{ESeraj2}.

\item The program \texttt{Test\_StstclFrmwrk\_Example\_LowAmp.m} is the implementation of an example showing the effects of calculating phase and frequency in low-amplitude analytic signal, as represented in \cite{ESeraj2}.

\item The program \texttt{Test\_StstclFrmwrk\_ROCCurves.m} calculates and illustrates the probability of signal detectability versus required SNR by using the probability of false alarm (as represented in \cite{ESeraj2} and Ch.6 of \cite{Richards}).
\end{itemize}

\subsubsection{Phase Related Quantities}
\label{subsubsec:phasefeatures}
The most popular phase related quantities, particularly used in brain-studies and neuroscience, are: Phase Derivative (PD), Phase Shift (PS), Phase Lock (PL), Phase Resetting (PR), Instantaneous Frequency and Phase-locking Value (PLV). In this toolbox, functions \texttt{Phase\_Features.m}, \texttt{Phase\_Features\_SingleCh.m} and \texttt{Phase\_Features\_MultiCh.m} are provided to calculate the first five features in single or multi-channel signals and the functions \texttt{PLV\_RawSig.m} and \texttt{PLV\_PhaseSeq.m} are generated to measure the PLV in multi-channel signals (multi-channel EEG records) or between separate signals. The underlying principals are stated in the followings. These scripts are the main software utilized for research studies presented in \cite{ESeraj5, ESeraj6, ESeraj7} and  \cite{ESeraj8}.

\paragraph{Phase Derivative (PD) and Instantaneous Frequency (IF):}
\label{parag:PD}
The PD is one of most common and well-known phase related quantities used in analyzing the phase of signals. In this toolbox, it can be calculated both in single and multi-channel modes through the functions \texttt{Phase\_Features\_SingleCh.m} or \texttt{Phase\_Features\_MultiCh.m}. In single channel case, for a signal $ x(t) $, the $ PD_s $ is computed as first order time derivative of the phase sequence, which is equal to the instantaneous frequency (IF) of the signal \cite{ESeraj2}.
\begin{equation}
PD_s(t) = \frac{d\phi_x(t)}{dt}
\label{eq:PDsingle}
\end{equation}
where the $ \phi_x(t) $ is the instantaneous phase sequence of $ x(t) $ captured at frequency $ f $. In multi-channel case, $ PD_m $ is calculated as the phase difference between two signals $ x(t) $ and $ y(t) $ as below \cite{ESeraj2}:
\begin{equation}
PD_m(t) = \phi_x(t) - \phi_y(t)
\label{eq:PDmulti}
\end{equation}

\paragraph{Phase Shift (PS), Phase Lock (PL) and Phase Resetting (PR) Events:}
\label{parag:PS}
Calculating the PS events consists of, first detecting the phase-displacements or phase-jumps, and then choosing the most significant ones as the PS events \cite{Thatcher3, WJFreeman3}. In other words, for computing the phase shift, the first step is to calculate the $ d(t) $ as,
\begin{equation}
d(t) = \frac{d(PD(t))}{dt}
\label{eq:d(t)}
\end{equation}
and then the PS can be obtained as follows \cite{ESeraj2}:
\begin{equation}
\forall t : PS = d(t)~~|~~d(t) \geq Th
\label{eq:PS}
\end{equation}
where $ Th $ is the threshold used to discriminate between significant and nonsignificant displacements or jumps. 

On the contrary, In the case of spontaneous or ongoing cerebral signals where there is no evoking stimulus, a near zero first derivative of PD demonstrates phase lock \cite{Thatcher3, WJFreeman3}. In other words, PL has an opposite definition to PS. Therefore, after calculating the $ d(t) $ as represented in equation~\ref{eq:d(t)}, the PL can be computed as follows:
\begin{equation}
\forall t : PL = d(t)~~|~~d(t) \leq Th
\label{eq:PL}
\end{equation}

Basically, phase resetting is made up of a phase shift followed by a phase difference stability, i.e. phase lock \cite{Pikovsky}. Thus, each pair of PS and PL, starting from the beginning of the PS and finishing by the end of PL (beginning of the next PS), is called a phase reset.

For calculating the PD, IF, PS, PL or PR, based on single or multi-channel phase features, you can either use the test m-files of the introduced functions, namely, 
\begin{center}
\texttt{Test\_Phase\_Features.m}\\
\texttt{Test\_Phase\_Features\_SingleCh.m}\\ \texttt{Test\_Phase\_Features\_MultiCh.m}
\end{center}
 or simply type:
\begin{center}
\texttt{[PR, PS, PL, PDV, PD] = Phase\_Features(phase\_sig1,  fs, Th, phase\_sig2)}
\end{center}
or
\begin{center}
\texttt{[PR, PS, PL, PDV, PD] = Phase\_Features\_MultiCh(meth, sig1, sig2, fs, f0, WS, ndft, pertnum)}
\end{center}
for multi-channel PS and type:
\begin{center}
\texttt{[IF, PR, PS, PL] = Phase\_Features(phase\_sig1,  fs, Th)}
\end{center}
or
\begin{center}
\texttt{[IF, PR, PS, PL] = Phase\_Features\_SingleCh(meth, sig, fs, f0, WS, ndft, pertnum)}
\end{center}
for signle-channel case. The difference between function \texttt{Phase\_Features.m} and functions \texttt{Phase\_Features\_MultiCh.m} or \texttt{Phase\_Features\_SingleCh.m} is that function \texttt{Phase\_Features.m} takes phase sequences as input and calculates the quantities; however, functions \texttt{Phase\_Features\_MultiCh.m} and \texttt{Phase\_Features\_SingleCh.m} take raw signals (i.e. EEG) as inputs, compute the instantaneous phase sequences of input signals and then measure required phase features.

\paragraph{Phase-locking Value (PLV):}
\label{parag:PLV}
PLV is a measure for quantifying how constant the phase difference between two signals is. In order to calculate the PLV for two signals (or channels) $x(t)$ and $y(t)$, the following steps are required \cite{JPLachaux}:
\begin{itemize}
\item Using narrow-band filters centered at $f$, calculate the instantaneous frequency-specific phase values $\phi_x(t,f)$ and $\phi_y(t,f)$.
\item Calculate the instantaneous phase-difference between $x(t)$ and $y(t)$ and quantify the local stability of this phase-difference over time:
\begin{equation}
\label{eq09}
PLV(f)=\left|\frac{1}{T}\sum_{t=1}^{T}\exp\left(j[\phi_y(t,f)-\phi_x(t,f)]\right)\right|
\end{equation}
where $T$ is the signal length and the summation is over all temporal samples of the instantaneous phases.
\end{itemize}
PLV varies between 0 and 1, corresponding to completely non-synchronized signals and  complete synchronization, respectively \cite{JPLachaux}. In this toolbox, provided functions calculate the pairwise PLV which is the calculated PLV value between each possible pair of signal channels in a desired frequency $ f $.

In the current toolbox, the functions \texttt{PLV\_RawSig.m} and \texttt{PLV\_PhaseSeq.m} are provided for this purpose. To operate these functions, you can either use the test m-files \texttt{Test\_PLV\_RawSig.m} and \texttt{Test\_PLV\_PhaseSeq.m}, or just simply type:
\begin{center}
\texttt{PLV = PLV\_RawSig(meth, sig1, fs, f0, WS, ndft, pertnum, sig2)}
\end{center}
for calculating PLV from raw signals (i.e. EEG signal) and type:
\begin{center}
\texttt{PLV = PLV\_PhaseSeq(phase\_sig1, phase\_sig2, phase\_sig3, ...)}
\end{center}
for measuring the pairwise PLV from phase sequences, where the last case can also be used in another form as:
\begin{center}
\texttt{PLV = PLV\_PhaseSeq(phase\_sig)}
\end{center}
where in first case \texttt{phase\_sig1, phase\_sig2, phase\_sig3, ...} are phase vectors; however \texttt{phase\_sig} in last case is a phase matrix with each row stating a phase sequence.

\section{Reference Manual}
\label{sec:referencemanual}

\subsection{\texttt{Phase\_Ex\_Trad.m}}
\subsection*{\fbox{\parbox{14.7cm}{\texttt{Phase\_Ex\_Trad.m}}}}
\paragraph*{Purpose:}
Instantaneous Phase (IP) estimation using the conventional analytic representation approach through FIR filtering and Hilbert Transform.
 
\paragraph*{Synopsis (global mode):}
\begin{center}
{\tt
 [phase, inst\_freq, amp] = Phase\_Ex\_Trad(x1, fs)
}
\end{center}

\paragraph*{Synopsis (local mode):}
\begin{center}
{\tt
 [phase, inst\_freq, amp] = Phase\_Ex\_Trad(x1, fs, WS, max\_f, ndft)
}
\end{center}

\paragraph*{Inputs:}
\begin{center}
\begin{tabular}{ll}
Input              & Description \\
\hline
\hline
\texttt{x1} & input signal \\
\texttt{fs}            & sampling frequency (Hz) \\
\hline
\texttt{WS}           & FIR filter's stop-band frequency (Hz) \\
\texttt{max\_f}     & maximum frequency (Hz) required to extract phase information\\
\texttt{ndft}     & number of frequency bins \\
\end{tabular}
\end{center}

\paragraph*{Defaults:}
\begin{center}
\begin{tabular}{lc}
Input              & Default Values \\
\hline
\hline
\texttt{WS}           & 1(Hz) \\
\texttt{max\_f}     & 30(Hz)\\
\texttt{ndft}     & 100 \\
\end{tabular}
\end{center}

\paragraph*{Outputs:}
\begin{center}
\begin{tabular}{ll}
Output              & Description \\
\hline
\hline
\texttt{phase} & Instantaneous Phase matrix \\
\texttt{inst\_freq}            & Instantaneous Frequency matrix \\
\texttt{amp}           & Instantaneous Amplitude matrix \\
\end{tabular}
\end{center}

\paragraph*{Notes:}
\begin{itemize}
\item The bandwidth \texttt{BW} in utilized FIR filter is approximately equal to $ \frac{\texttt{WS}}{2} $.

\item The outputs are calculated as matrices for the frequency interval [1 \texttt{max\_f}] where the rows include phase information regarding the corresponding frequency components (for example 10th row in output phase matrix includes instantaneous phase of 10(Hz) estimated from input signal).

\item While specifying a value to one of the parameters having default values, an empty bracket [] must be used for non-specified parameters. If you're using the global synopsis, empty bracket is not required.

\item The m-file \texttt{Test\_Phase\_Ex\_Trad.m} is a demo program to operate this function.
\end{itemize}

\subsection{\texttt{Phase\_Ex\_ZPPP.m}}
\subsection*{\fbox{\parbox{14.7cm}{\texttt{Phase\_Ex\_ZPPP.m}}}}
\paragraph*{Purpose:}
Instantaneous Phase (IP) estimation using the Zer-pole Perturbation Phase estimation method (Z-PPP) \cite{ESeraj1} through analytic representation IIR filters and forward-backward filtering. Refer to \cite{ESeraj1} for detailed description.
 
\paragraph*{Synopsis (global mode):}
\begin{center}
{\tt
 [phase, inst\_freq, amp] = Phase\_Ex\_ZPPP(x1, fs)
}
\end{center}

\paragraph*{Synopsis (local mode):}
\begin{center}
{\tt
 [phase, inst\_freq, amp] = Phase\_Ex\_ZPPP(x1, fs, WS, max\_f, ndft, pertnum)
}
\end{center}

\paragraph*{Inputs:}
\begin{center}
\begin{tabular}{ll}
Input              & Description \\
\hline
\hline
\texttt{x1} & input signal \\
\texttt{fs}            & sampling frequency (Hz) \\
\hline
\texttt{WS}           & IIR filter's stop-band frequency (Hz) \\
\texttt{max\_f}     & maximum frequency (Hz) required to extract phase information\\
\texttt{ndft}     & number of frequency bins \\
\texttt{pertnum}     & number of attempts for perturbing filter's zeros and poles \\
\end{tabular}
\end{center}

\paragraph*{Defaults:}
\begin{center}
\begin{tabular}{lc}
Input              & Default Values \\
\hline
\hline
\texttt{WS}           & 1(Hz) \\
\texttt{max\_f}     & 30(Hz)\\
\texttt{ndft}     & 100 \\
\texttt{pertnum}  & 100 \\
\end{tabular}
\end{center}

\paragraph*{Outputs:}
\begin{center}
\begin{tabular}{ll}
Output              & Description \\
\hline
\hline
\texttt{phase} & Instantaneous Phase matrix \\
\texttt{inst\_freq}            & Instantaneous Frequency matrix \\
\texttt{amp}           & Instantaneous Amplitude matrix \\
\end{tabular}
\end{center}

\paragraph*{Notes:}
\begin{itemize}
\item The bandwidth \texttt{BW} in utilized IIR filter is equal to $ \frac{\texttt{WS}}{2} $.

\item The outputs are calculated as matrices for the frequency interval [1 \texttt{max\_f}] where the rows include phase information regarding the corresponding frequency components (for example 10th row in output phase matrix includes instantaneous phase of 10(Hz) estimated from input signal).

\item While specifying a value to one of the parameters having default values, an empty bracket [] must be used for non-specified parameters. If you're using the global synopsis, empty bracket is not required.

\item The m-file \texttt{Test\_Phase\_Ex\_ZPPP.m} is a demo program to operate this function.
\end{itemize}

\subsection{\texttt{Phase\_Ex\_TFP.m}}
\subsection*{\fbox{\parbox{14.7cm}{\texttt{Phase\_Ex\_TFP.m}}}}
\paragraph*{Purpose:}
Instantaneous Phase estimation using the Transfer-Function Perturbation Phase estimation method (TFP) \cite{ESeraj2} through analytic representation, IIR filters and forward-backward filtering. Refer to the \cite{ESeraj2} for detailed description.
 
\paragraph*{Synopsis (global mode):}
\begin{center}
{\tt
 [phase, inst\_freq, amp] = Phase\_Ex\_TFP(x1, fs)
}
\end{center}

\paragraph*{Synopsis (local mode):}
\begin{center}
{\tt
 [phase, inst\_freq, amp] = Phase\_Ex\_TFP(x1, fs, f0, bw\_base, f0\_dev, bw\_base\_dev, dither\_std, pertnum)
}
\end{center}

\paragraph*{Inputs:}
\begin{center}
\begin{tabular}{ll}
Input              & Description \\
\hline
\hline
\texttt{x1} & input signal \\
\texttt{fs}            & sampling frequency (Hz) \\
\hline
\texttt{f0}           & center frequency of the frequency filter (Hz) \\
\texttt{bw\_base}     & bandwidth of the frequency filter (Hz) \\
\texttt{f0\_dev}     & center frequency deviation range (Hz) \\
\texttt{bw\_base\_dev}     & bandwidth deviation range (Hz) \\
\texttt{dither\_std}     & dither noise level \\
\texttt{pertnum}     & number of attempts for perturbing filter's transfer function \\
\end{tabular}
\end{center}

\paragraph*{Defaults:}
\begin{center}
\begin{tabular}{lc}
Input              & Default Values \\
\hline
\hline
\texttt{f0}           & 10 (Hz) *** for EEG alpha rhythms \\
\texttt{bw\_base}     & 4 (Hz) *** for EEG alpha rhythms \\
\texttt{f0\_dev}     & 1e-6 (Hz) \\
\texttt{bw\_base\_dev}     & 1e-1 (Hz) \\
\texttt{dither\_std}     & 1e-4 \\
\texttt{pertnum}     & 100 \\
\end{tabular}
\end{center}

\paragraph*{Outputs:}
\begin{center}
\begin{tabular}{ll}
Output              & Description \\
\hline
\hline
\texttt{phase} & Instantaneous Phase matrix \\
\texttt{inst\_freq}            & Instantaneous Frequency matrix \\
\texttt{amp}           & Instantaneous Amplitude matrix \\
\end{tabular}
\end{center}

\paragraph*{Notes:}
\begin{itemize}
\item By default, without specifying the frequency band of interest for phase estimation through TFP method, this function calculates the instantaneous parameters, i.e. IP, IF and IE, for the alpha rhythms (8-12Hz) of input EEG signal.

\item While specifying a value to one of the parameters having default values, an empty bracket [] must be used for non-specified parameters. If you're using the global synopsis, empty bracket is not required.

\item The m-file \texttt{Test\_Phase\_Ex\_TFP.m} is a demo program to operate this function.
\end{itemize}

\subsection{\texttt{Phase\_Features.m}}
\subsection*{\fbox{\parbox{14.7cm}{\texttt{Phase\_Features.m}}}}
\paragraph*{Purpose:}
Calculating popular phase related quantities, i.e. Phase Shift (PS), Phase Lock (PL), Phase Reset (PR), Phase Difference (PD) and Instantaneous Frequency (IF) in SINGLE or MULTI-Channel modes using phase sequences.
 
\paragraph*{Synopsis (Multi-channel):}
\begin{center}
{\tt
 [PR, PS, PL, PDV, PD] = Phase\_Features(phase\_sig1,  fs, Th)
 
 [PR, PS, PL, PDV, PD] = Phase\_Features(phase\_sig1,  fs, Th, phase\_sig2)
}
\end{center}

\paragraph*{Synopsis (Single-channel):}
\begin{center}
{\tt
 [IF, PR, PS, PL] = Phase\_Features(phase\_sig1,  fs, Th)
}
\end{center}

\paragraph*{Inputs:}
\begin{center}
\begin{tabular}{ll}
Input              & Description \\
\hline
\hline
\texttt{phase\_sig1} & instantaneous phase vector or matrix \\
\texttt{fs}            & sampling frequency (Hz) \\
\texttt{Th}           & threshold value used to detect phase shift events (Radians) \\
\hline
\texttt{phase\_sig2}     & instantaneous phase vector\\
\end{tabular}
\end{center}

\paragraph*{Outputs:}
\begin{center}
\begin{tabular}{ll}
Output              & Description \\
\hline
\hline
\texttt{PR} & Phase Resetting events vector \\
\texttt{PS}            & Phase Shift events vector \\
\texttt{PL}           & Phase Lock events vector \\
\texttt{IF}           & Instantaneous Frequency vector \\
\texttt{PD}           & Phase Difference vector \\
\texttt{PDV}           & Phase Difference Variations (first order time derivative) vector \\
\end{tabular}
\end{center}

\paragraph*{Notes:}
\begin{itemize}

\item The option \texttt{phase\_sig2} is provided in case that someone needs to calculate the phase features between two separate signals; however, the recommended option is to use the input \texttt{phase\_sig1} as a phase matrix (with size 2*N) for calculating phase quantities in multi-channel case.

\item In single-channel mode, the input \texttt{phase\_sig1} have to be a vector and in multi-channel mode, it could be a vector or a matrix, as stated above.

\item The m-file \texttt{Test\_Phase\_Features.m} is a demo program to operate this function.
\end{itemize}

\subsection{\texttt{Phase\_Features\_MultiCh.m}}
\subsection*{\fbox{\parbox{14.7cm}{\texttt{Phase\_Features\_MultiCh.m}}}}
\paragraph*{Purpose:}
Calculating popular phase related quantities, i.e. Phase Shift (PS), Phase Lock (PL), Phase Reset (PR), Phase Difference (PD) and Phase Difference Derivatives (PDV) in MULTI-Channel mode.
 
\paragraph*{Synopsis (global mode):}
\begin{center}
{\tt
 [PR, PS, PL, PDV, PD] = Phase\_Features\_MultiCh(meth, sig1, sig2, fs, f0)
}
\end{center}

\paragraph*{Synopsis (local mode):}
\begin{center}
{\tt
 [PR, PS, PL, PDV, PD] = Phase\_Features\_MultiCh(meth, sig1, sig2, fs, f0, WS, ndft, pertnum)
}
\end{center}

\paragraph*{Inputs:}
\begin{center}
\begin{tabular}{ll}
Input              & Description \\
\hline
\hline
\texttt{meth} & character specifying utilized method for phase estimation \\
\texttt{sig1} & input raw signal \#1 vector \\
\texttt{sig2} & input raw signal \#2 vector\\
\texttt{fs}            & sampling frequency (Hz) \\
\texttt{f0}            & frequency of interest (Hz) \\
\hline
\texttt{WS}           & utilized filter's stop-band frequency (Hz) \\
\texttt{ndft}     & number of frequency bins \\
\texttt{pertnum}     & number of attempts for perturbing filter's zeros and poles \\
\end{tabular}
\end{center}

\paragraph*{Defaults:}
\begin{center}
\begin{tabular}{lc}
Input              & Default Values \\
\hline
\hline
\texttt{WS}           & 1(Hz) \\
\texttt{ndft}     & 100 \\
\texttt{pertnum}  & 100 \\
\end{tabular}
\end{center}

\paragraph*{Outputs:}
\begin{center}
\begin{tabular}{ll}
Output              & Description \\
\hline
\hline
\texttt{PR} & Phase Resetting events vector \\
\texttt{PS}            & Phase Shift events vector \\
\texttt{PL}           & Phase Lock events vector \\
\texttt{PD}           & Phase Difference vector \\
\texttt{PDV}           & Phase Difference Variations (first order time derivative) vector \\
\end{tabular}
\end{center}

\paragraph*{Notes:}
\begin{itemize}
\item The bandwidth \texttt{BW} in utilized filter, i.e. IIR or FIR, is equal to $ \frac{\texttt{WS}}{2} $.

\item The available values for the input \texttt{meth} are either 'ZPPP' or 'Trad' where the first uses the \texttt{Phase\_Ex\_ZPPP} and the other one uses the \texttt{Phase\_Ex\_Trad} for estimating initial phase sequence.

\item While specifying a value to one of the parameters having default values, an empty bracket [] must be used for non-specified parameters. If you're using the global synopsis, empty bracket is not required.

\item The m-file \texttt{Test\_Phase\_Features\_MultiCh.m} is a demo program to operate this function.
\end{itemize}

\subsection{\texttt{Phase\_Features\_SingleCh.m}}
\subsection*{\fbox{\parbox{14.7cm}{\texttt{Phase\_Features\_SingleCh.m}}}}
\paragraph*{Purpose:}
Calculating popular phase related quantities, i.e. Phase Shift (PS), Phase Lock (PL), Phase Reset (PR) and Instantaneous Frequency (IF) in SINGLE-Channel mode.
 
\paragraph*{Synopsis (global mode):}
\begin{center}
{\tt
 [IF, PR, PS, PL] = Phase\_Features\_SingleCh(meth, sig, fs, f0)
}
\end{center}

\paragraph*{Synopsis (local mode):}
\begin{center}
{\tt
 [IF, PR, PS, PL] = Phase\_Features\_SingleCh(meth, sig, fs, f0, WS, ndft, pertnum)
}
\end{center}

\paragraph*{Inputs:}
\begin{center}
\begin{tabular}{ll}
Input              & Description \\
\hline
\hline
\texttt{meth} & character specifying utilized method for phase estimation \\
\texttt{sig} & input raw signal vector \\
\texttt{fs}            & sampling frequency (Hz) \\
\texttt{f0}            & frequency of interest (Hz) \\
\hline
\texttt{WS}           & utilized filter's stop-band frequency (Hz) \\
\texttt{ndft}     & number of frequency bins \\
\texttt{pertnum}     & number of attempts for perturbing filter's zeros and poles \\
\end{tabular}
\end{center}

\paragraph*{Defaults:}
\begin{center}
\begin{tabular}{lc}
Input              & Default Values \\
\hline
\hline
\texttt{WS}           & 1(Hz) \\
\texttt{ndft}     & 100 \\
\texttt{pertnum}  & 100 \\
\end{tabular}
\end{center}

\paragraph*{Outputs:}
\begin{center}
\begin{tabular}{ll}
Output              & Description \\
\hline
\hline
\texttt{PR} & Phase Resetting events vector \\
\texttt{PS}            & Phase Shift events vector \\
\texttt{PL}           & Phase Lock events vector \\
\texttt{IF}           & Instantaneous Frequency vector \\
\end{tabular}
\end{center}

\paragraph*{Notes:}
\begin{itemize}
\item The bandwidth \texttt{BW} in utilized filter, i.e. IIR or FIR, is equal to $ \frac{\texttt{WS}}{2} $.

\item The available values for the input \texttt{meth} are either 'ZPPP' or 'Trad' where the first uses the \texttt{Phase\_Ex\_ZPPP} and the other one uses the \texttt{Phase\_Ex\_Trad} for estimating initial phase sequence.

\item While specifying a value to one of the parameters having default values, an empty bracket [] must be used for non-specified parameters. If you're using the global synopsis, empty bracket is not required.

\item The m-file \texttt{Test\_Phase\_Features\_SingleCh.m} is a demo program to operate this function.
\end{itemize}

\subsection{\texttt{PLV\_PhaseSeq.m}}
\subsection*{\fbox{\parbox{14.7cm}{\texttt{PLV\_PhaseSeq.m}}}}
\paragraph*{Purpose:}
Calculating Phase Locking Value (PLV) matrix (Pairwise PLV) using phase sequences.
 
\paragraph*{Synopsis:}
\begin{center}
{\tt
 PLV = PLV\_PhaseSeq(phase\_sig)
 
 PLV = PLV\_PhaseSeq(phase\_sig1, phase\_sig2, phase\_sig3, ...)
}
\end{center}

\paragraph*{Inputs:}
\begin{center}
\begin{tabular}{ll}
Input              & Description \\
\hline
\hline
\texttt{phase\_sig} & input phase matrix \\
\hline
\texttt{phase\_sig1}           & input phase vector \#1 \\
\texttt{phase\_sig2}     & input phase vector \#2 \\
$ \bullet $ & $ \bullet $ \\
$ \bullet $ & $ \bullet $ \\
\end{tabular}
\end{center}

\paragraph*{Outputs:}
\begin{center}
\begin{tabular}{ll}
Output              & Description \\
\hline
\hline
\texttt{PLV} & Pairwise PLV matrix \\
\end{tabular}
\end{center}

\paragraph*{Notes:}
\begin{itemize}
\item While using the first case, the \texttt{phase\_sig} have to be a matrix with at least two rows where each row represents a phase signal. In second case, each of the \texttt{phase\_sig1...phase\_sign} are row vectors of phase sequences. This option is provided in case that someone needs to calculate the PLV matrix between separate phase signals.

\item In case that you need a single PLV value between signals (and not a pairwise PLV matrix), you can just simply use the non-diagonal values in output matrix.

\item The m-file \texttt{Test\_PLV\_PhaseSeq.m} is a demo program to operate this function.
\end{itemize}

\subsection{\texttt{PLV\_RawSig.m}}
\subsection*{\fbox{\parbox{14.7cm}{\texttt{PLV\_RawSig.m}}}}
\paragraph*{Purpose:}
Calculating Phase Locking Value (PLV) matrix (Pairwise) using raw signals.
 
\paragraph*{Synopsis (global mode):}
\begin{center}
{\tt
 PLV = PLV\_RawSig(meth, sig1, fs, f0)
}
\end{center}

\paragraph*{Synopsis (local mode):}
\begin{center}
{\tt
 PLV = PLV\_RawSig(meth, sig1, fs, f0, WS, ndft, pertnum, sig2)
}
\end{center}

\paragraph*{Inputs:}
\begin{center}
\begin{tabular}{ll}
Input              & Description \\
\hline
\hline
\texttt{meth} & character specifying utilized method for phase estimation \\
\texttt{sig1} & input raw signal \#1 vector \\
\texttt{fs}            & sampling frequency (Hz) \\
\texttt{f0}            & frequency of interest (Hz) \\
\hline
\texttt{WS}           & utilized filter's stop-band frequency (Hz) \\
\texttt{ndft}     & number of frequency bins \\
\texttt{pertnum}     & number of attempts for perturbing filter's zeros and poles \\
\texttt{sig2} & input raw signal \#2 vector \\
\end{tabular}
\end{center}

\paragraph*{Defaults:}
\begin{center}
\begin{tabular}{lc}
Input              & Default Values \\
\hline
\hline
\texttt{WS}           & 1(Hz) \\
\texttt{ndft}     & 100 \\
\texttt{pertnum}  & 100 \\
\end{tabular}
\end{center}

\paragraph*{Outputs:}
\begin{center}
\begin{tabular}{ll}
Output              & Description \\
\hline
\hline
\texttt{PLV} & Pairwise PLV matrix \\
\end{tabular}
\end{center}

\paragraph*{Notes:}
\begin{itemize}
\item The bandwidth \texttt{BW} in utilized filter, i.e. IIR or FIR, is equal to $ \frac{\texttt{WS}}{2} $.

\item The available values for the input \texttt{meth} are either 'ZPPP' or 'Trad' where the first uses the \texttt{Phase\_Ex\_ZPPP} and the other one uses the \texttt{Phase\_Ex\_Trad} for estimating initial phase sequence.

\item While specifying a value to one of the parameters having default values, an empty bracket [] must be used for non-specified parameters. If you're using the global synopsis, empty bracket is not required.

\item In case that you need a single PLV value between signals (and not a pairwise PLV matrix), you can just simply use the non-diagonal values in output matrix.

\item The option \texttt{sig2} is provided in case that someone needs to calculate the PLV value between separate raw signals. To have the PLV value, you can just simply use one of the non-diagonal values in output $ 2*2 $ matrix. 

\item The m-file \texttt{Test\_PLV\_RawSig.m} is a demo program to operate this function.
\end{itemize}

\subsection{\texttt{Synth\_EEG.m}}
\subsection*{\fbox{\parbox{14.7cm}{\texttt{Synth\_EEG.m}}}}
\paragraph*{Purpose:}
Generating synthetic EEG signal with spectral characteristics similar to a real EEG, using Autoregressive (AR) model and innovation filter. Could be used in cross-validations with real EEG signal.
 
\paragraph*{Synopsis (global mode):}
\begin{center}
{\tt
eeg\_synth = Synth\_EEG(sig, fs)
}
\end{center}

\paragraph*{Synopsis (local mode):}
\begin{center}
{\tt
eeg\_synth = Synth\_EEG(sig, fs, duration, win, AR\_ord)
}
\end{center}

\paragraph*{Inputs:}
\begin{center}
\begin{tabular}{ll}
Input              & Description \\
\hline
\hline
\texttt{sig} & input raw EEG signal vector \\
\texttt{fs}            & sampling frequency of input signal (Hz) \\
\hline
\texttt{duration}           & total required signal duration (Seconds) \\
\texttt{win}     & temporal window length to have a stationary signal (Seconds) \\
\texttt{AR\_ord}     & order of AR model \\
\end{tabular}
\end{center}

\paragraph*{Defaults:}
\begin{center}
\begin{tabular}{lc}
Input              & Default Values \\
\hline
\hline
\texttt{duration}           & length of input raw EEG signal \\
\texttt{win}     & 3 or 4 or 5 sec (depends on input EEG signal) \\
\texttt{AR\_ord}  & 20 \\
\end{tabular}
\end{center}

\paragraph*{Outputs:}
\begin{center}
\begin{tabular}{ll}
Output              & Description \\
\hline
\hline
\texttt{eeg\_synth} & Synthetic EEG signal \\
\end{tabular}
\end{center}

\paragraph*{Notes:}
\begin{itemize}
\item While specifying a value to one of the parameters having default values, an empty bracket [] must be used for non-specified parameters. If you're using the global synopsis, empty bracket is not required.

\item The m-file \texttt{Test\_Synth\_EEG.m} is a demo program to operate this function.
\end{itemize}

\section{Outcomes}
\label{sec:outcomes}
Below are presented some example figures generated by the test m-files provided within the toolbox. The following figures are generated using the Z-PPP and TFP phase estimation methods \cite{ESeraj1, ESeraj2}. Similar forms of outcomes can be observed using the traditional method.
\begin{figure}[th]
\centering
\includegraphics[scale=0.5]{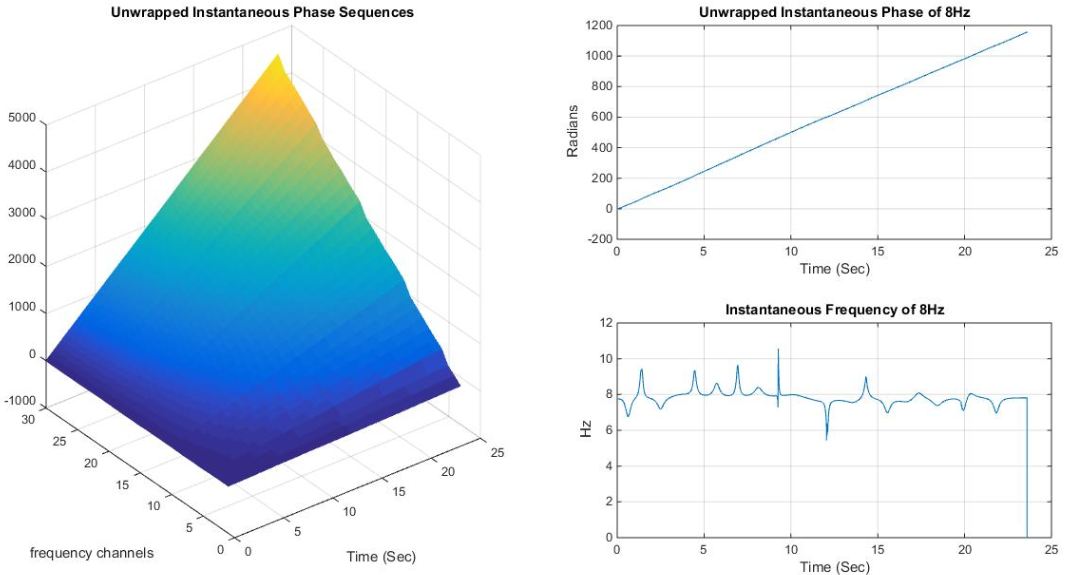}
\caption{Instantaneous Phase (IP) sequences of [1 30]Hz (left panel). IP and IF of $ 8 $Hz, (right panel, top and bottom figures respectively). }
\label{fig:01}
\end{figure}
\begin{figure}[th]
\centering
\includegraphics[scale=0.5]{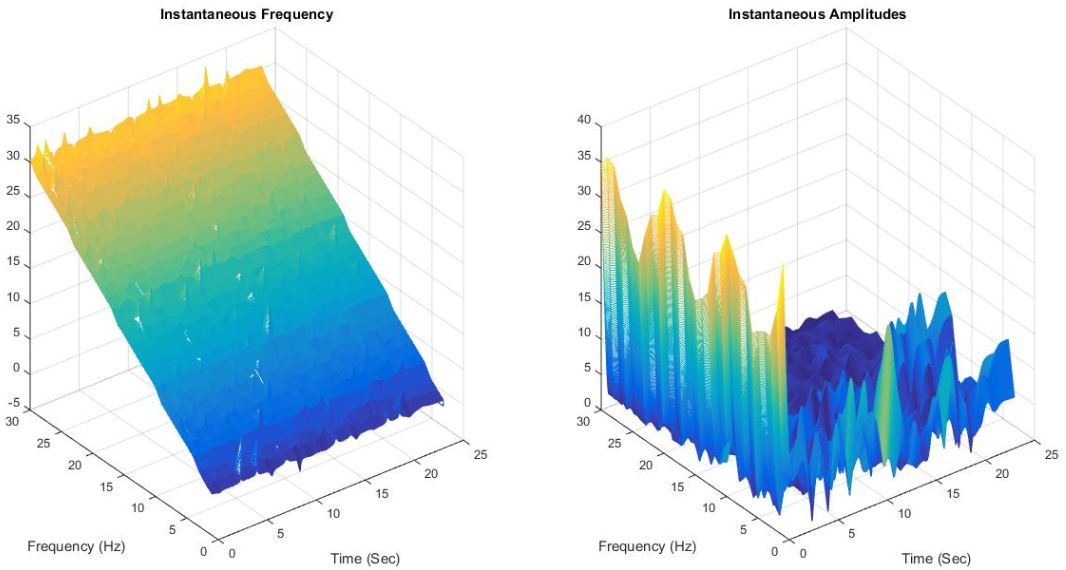}
\caption{Instantaneous Frequency (IF) and Instantaneous Amplitude (IA) sequences of [1 30]Hz, left and right side panels respectively.}
\label{fig:02}
\end{figure}
\begin{figure}[th]
\centering
\includegraphics[scale=0.55]{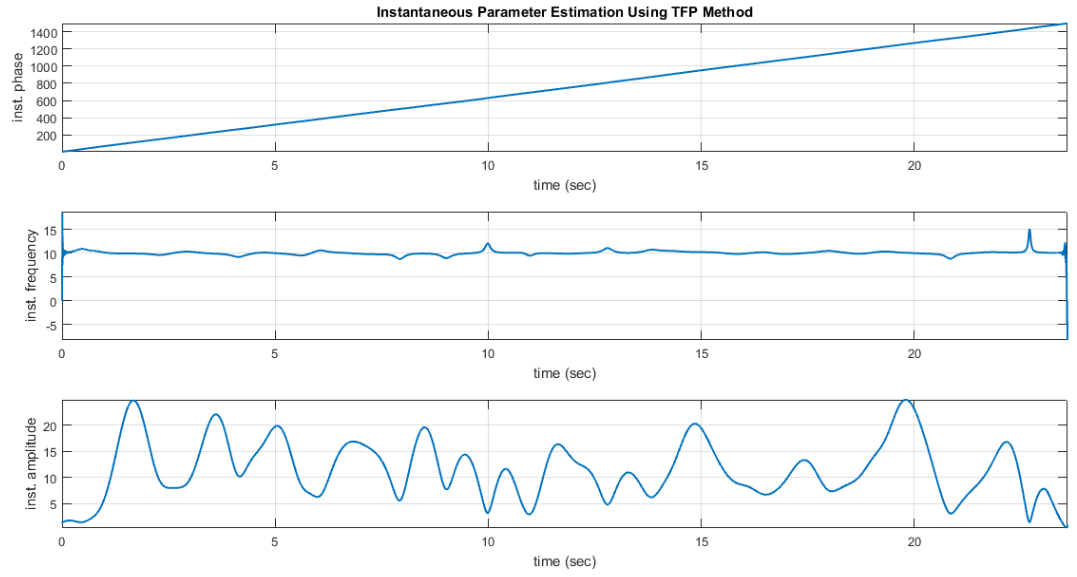}
\caption{Instantaneous Phase (IP), Instantaneous Frequency (IF) and Instantaneous Amplitude (IA) sequences of alpha rhythms [8 12]Hz of a sample EEG signal, from top to bottom respectively.}
\label{fig:07}
\end{figure}
\begin{figure}[th]
\centering
\includegraphics[scale=0.45]{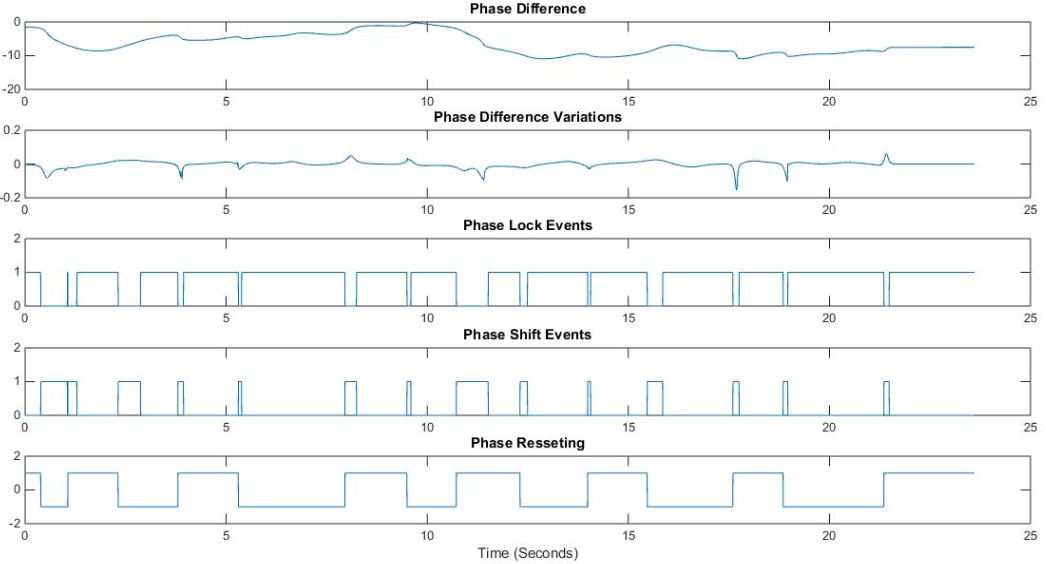}
\caption{Phase related quantities, MULTI-channel mode.}
\label{fig:03}
\end{figure}
\begin{figure}[th]
\centering
\includegraphics[scale=0.45]{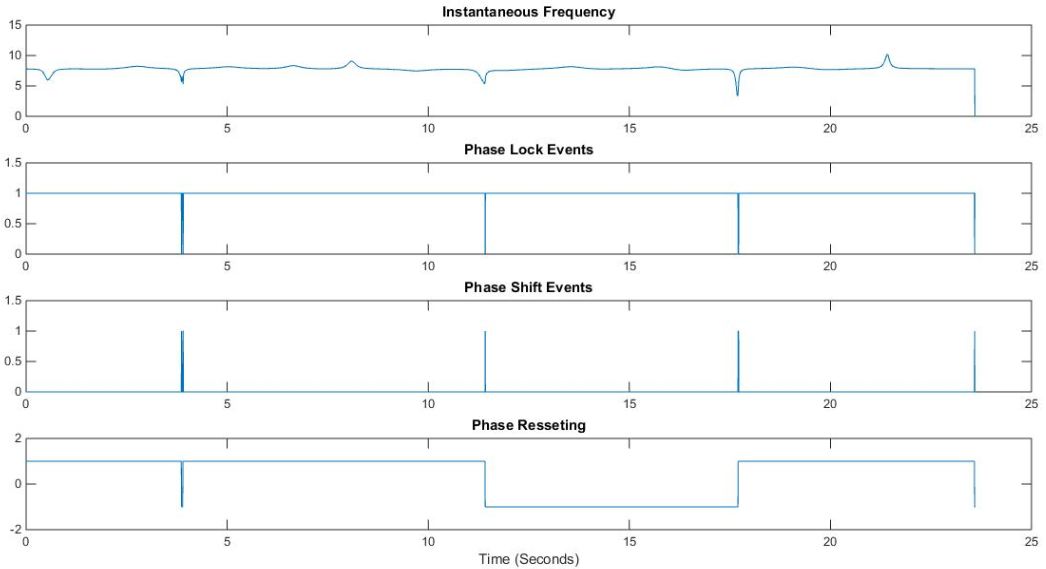}
\caption{Phase related quantities, SINGLE-channel mode.}
\label{fig:04}
\end{figure}
\begin{figure}[th]
\centering
\includegraphics[scale=0.45]{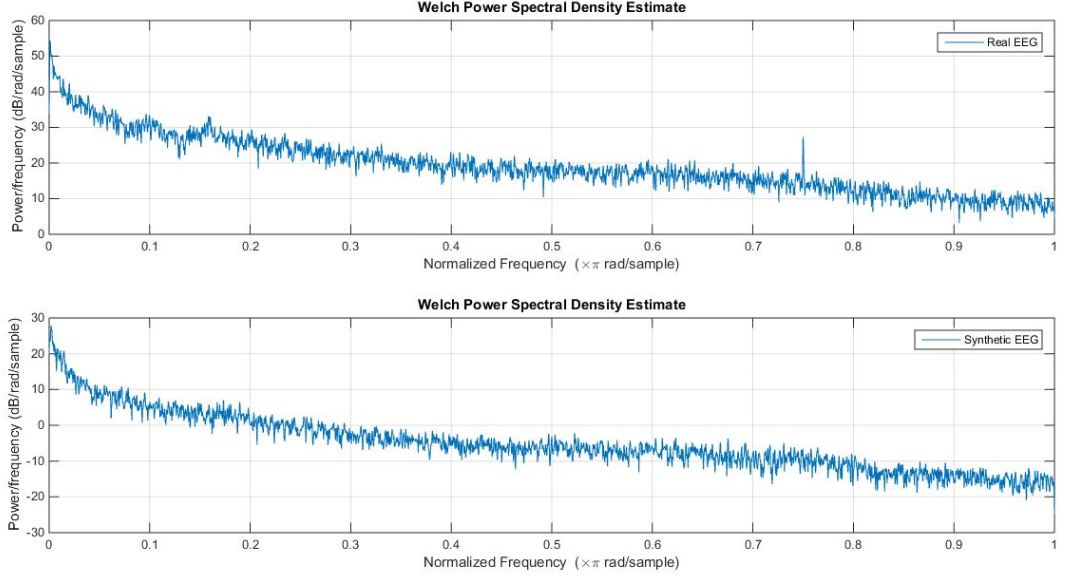}
\caption{Welch power spectral density (PSD) of real and synthetic EEG signals.}
\label{fig:06}
\end{figure}
\begin{figure}[th]
\centering
\includegraphics[scale=0.5]{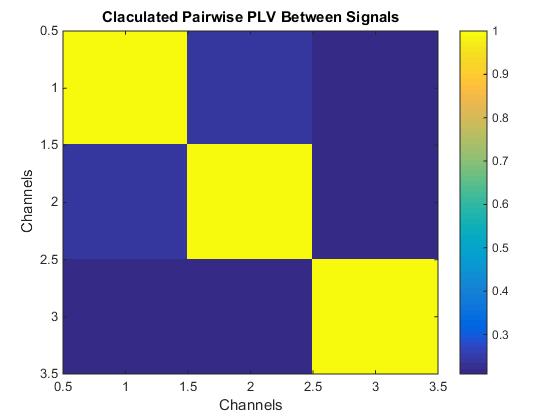}
\caption{Pairwise PLV matrix calculated between three sample EEG signals in dataset.}
\label{fig:05}
\end{figure}


\newpage
\section{References}
\begin{thebibliography}{99}
\bibitem{ESeraj1} Esmaeil Seraj and Reza Sameni, “Robust Electroencephalogram Phase Estimation with Applications in Brain-computer Interface Systems,”Physiological Measurement, vol. 38, no. 3, pp. 501–523, Feb. 2017. [Published Online] DOI: 10.1088/1361-6579/aa5bba

\bibitem{ESeraj2} Reza Sameni and Esmaeil Seraj, “A Robust Statistical Framework for Instantaneous Electroencephalogram Phase and Frequency Analysis,” Physiological Measurement, vol. 38, no. 12, pp. 2141–2163, Nov. 2017. [Published Online] DOI: 10.1088/1361-6579/aa93a1

\bibitem{ESeraj3} Esmaeil Seraj, “Cerebral Signal Phase Analysis Toolbox - User Guide Version 2.3,” arXiv Preprint, Dec. 2017 [Online]. Available: \underline{https://arxiv.org/abs/1610.02249}

\bibitem{ESeraj5} Foroozan Karimzadeh, Reza Boostani, Esmaeil Seraj, Reza Sameni, “A distributed classification procedure for automatic sleep stage scoring based on instantaneous electroencephalogram phase and envelope features,” IEEE Transactions on Neural Systems \& Rehabilitation Engineering, vol. 26, no. 2, pp. 362-370, Feb. 2018. DOI: 10.1109/TNSRE.2017.2775058

\bibitem{ESeraj6} Esmaeil Seraj, “An Investigation on the Utility and Reliability of Electroencephalogram Phase Signal upon Interpreting Cognitive Responses in the Brain: A Critical Discussion,” Journal of Advanced Medical Sciences and Applied Technologies, vol. 2, no. 4, pp. 299-312, Jan. 2017. [Published Online] DOI: http://dx.doi.org/10.18869\%2Fnrip.jamsat.2.4.299

\bibitem{ESeraj7} Esmaeil Seraj and Foroozan Karimzadeh, “Improved Detection Rate in Motor Imagery Based BCI Systems Using Combination of Robust Analytic Phase and Envelope Features,” 25th Iranian Conference on Electrical Engineering (ICEE), KNT University of Technology, Tehran, Iran, May 2017. [Published Online] DOI: 10.1109/IranianCEE.2017.7985458

\bibitem{ESeraj8} Esmaeil Seraj, “Cerebral Synchrony Assessment: A General Review on Cerebral Signals' Synchronization Estimation Concepts and Methods,” arXiv Preprint, Dec. 2016 [Online]. Available: https://arxiv.org/abs/1612.04295

\bibitem{RSameni} Reza Sameni, The Open-Source Electrophysiological Toolbox (OSET), [Online] version 3.1 (2014). URL \underline{http://www.oset.ir}

\bibitem{Andrzejak} R.~G.~Andrzejak, K.~Lehnertz, C.~Rieke, F.~Mormann, P.~David and C.~E.~Elger. \lq\lq Indications of nonlinear deterministic and finite dimensional structures in time series of brain electrical activity: Dependence on recording region and brain state,\rq\rq Phys. Rev., vol.~64, 2001.

\bibitem{MChavez} M.~Chavez, M.~Besserve, C.~Adam, J.~Martinerie, \lq\lq Towards a Proper Estimation of Phase Synchronization From Time Series, \rq\rq \emph{Journal of Neuroscience Methods}, vol.~154, 2006.

\bibitem{BPicinbono} B.~Picinbono, \lq\lq On the Instantaneous Amplitude and Phase of Signals, \rq\rq \emph{IEEE Transaction Signal Processesing}, vol.~45, pp.~552–560, 1997.

\bibitem{Thatcher3} R.~W.~Thatcher, D.~North and C.~Biver, \lq\lq Intelligence and EEG phase reset: A two compartmental model of phase shift and lock,\rq\rq \emph{Neuroimage}, vol.~42, pp.~1639-1653, 2008b.

\bibitem{WJFreeman3} W.~J.~Freeman, B.~C.~Burke and M.~D.~Homes, \lq\lq Aperiodic phase-resetting in scalp EEG of beta-gamma-oscillations by state transitions at alpha-thetarates,\rq\rq \emph{Human Brain Map.}, vol.~19, pp.~248-272, 2003.

\bibitem{Pikovsky} A.~Pikovsky, M.~G.~Rosenblum and J.~Kurths, \lq\lq Synchronization: A universal concept in nonlinear sciences,\rq\rq \emph{Cambridge Univ. Press, New York}, 2003.

\bibitem{JPLachaux} J.~P.~Lachaux, E.~Rodriguez, J.~Martinerie,F.~J.~Varela, \lq\lq Measuring Phase Synchrony in Brain Signals, \rq\rq \emph{Human Brain Mapping}, vol.~8, pp.~194–208, 1999.

\bibitem{Richards} M.~A.~Richards, \lq\lq Fundamentals of radar signal processing,\rq\rq Tata McGraw-Hill Education, 2005

\end{thebibliography}
\end{document}